\documentclass[3p,times,twocolumn]{elsarticle}

\usepackage{ecrc}

\volume{00}

\firstpage{1}

\journalname{Nuclear Physics B Proceedings Supplement}

\runauth{}

\jid{nuphbp}

\jnltitlelogo{Nuclear Physics B Proceedings Supplement}

\usepackage{amssymb}

\usepackage[figuresright]{rotating}
\usepackage{caption}

\newcommand{\be}{\begin{equation}}
\newcommand{\ee}{\end{equation}}
\newcommand{\ba}{\begin{eqnarray}}
\newcommand{\ea}{\end{eqnarray}}
\newcommand{\trc}[1]{\mathrm{tr}_c\left(#1\right)}
\newcommand{\trf}[1]{\mathrm{tr}_F\left(#1\right)}

\newcommand{\mphys}{M^2_{\mathrm{phys}}}
\newcommand{\nn}{\nonumber}

\begin{document}

\begin{frontmatter}



\dochead{}

\title{Effective Theories for QCD-like at TeV Scale}


\author[jl]{Jie Lu}
\fntext[jl]{Speaker}
\address{IFIC, Universitat de Valencia - CSIC, 
        Apt. Correus 22085, E-46071 Valencia, Spain
       }

\author{Johan Bijnens}
\address{Department of Astronomy and Theoretical Physics, Lund University,
S\`olvegatan 14A, SE 223-62 Lund, Sweden}
 
\begin{abstract}
We study the Effective Field Theory of three QCD-like theories, which can be classified by having quarks in a complex, real or pseudo-real representations of the gauge group. 
The Lagrangians are written in a very similar way so that the calculations can be done using techniques from Chiral Perturbation Theory (ChPT). 
We calculated the vacuum-expectation-value, the mass and the decay constant of pseudo-Goldstone Bosons up to next-to-next-to leading order (NNLO) \cite{Bijnens:2009qm}.  
The various channels of general $n$ flavour meson-meson scattering of the three theories are systematically studied and calculated up to NNLO \cite{Bijnens:2011fm}. 
We also calculated the vector, axial-vector, scalar, pseudo-scalar two-point functions and pseudo-scalar decay constant up NNLO order \cite{Bijnens:2011xt}. 
The analytic expressions of the S parameter for the three different QCD-like theories are obtained at TeV scale. Our results are useful for chiral extrapolation in lattice calculation on theory of strong dynamical and finite baryon density.
\end{abstract}


\end{frontmatter}


\section{Introduction}
Strong dynamical electroweak symmetry breaking (EWSB) is one important candidate theory for beyond Standard Model (SM). Although the SM-like Higgs Boson is discovered at LHC \cite{Aad:2012tfa, Chatrchyan:2012ufa}, it is still possible that it is composite, which arises from the pseudo-Goldstone Boson modes of new strong interaction at TeV scale, e.g., the Technicolor theory \cite{Technicolor1,Technicolor2} and other composite Higgs theories \cite{Bellazzini:2014yua}.

However, it is very difficult to use perturbative method in the strong interaction region. Lattice simulation is probably the most promising way for this problem. For computing the quantities in Technicolor theory, one has to push the calculation to the chiral limit, i.e., the massless quark limit, this is very time consuming and expensive \cite{lattice}. Therefore one can extrapolate the numerical data to the chiral limit using the analytic results from ChPT , which is called Chiral extrapolation. 

In this proceeding, we introduce the series of works on Effective Field Theory of three QCD-like theories, which are distinguished by having the (techni-)quarks live in a complex, real or pseudo-real representation of the gauge group. For $n$ flavours of identical quarks, this corresponds to the symmetry breaking pattern of $SU(n)_L\times SU(n)_R\to SU(n)_V$, $SU(2n)\to SO(2n)$ and $SU(2n)\to Sp(2n)$ respectively.  These theories can be used to characterize some Technicolor models with vector-like gauge bosons.  

QCD-like theories are also important in the theory of finite baryon density, where the normal QCD with chemical potential term suffers the sign problem in Lattice simulation. The real and pseudo-real case allow to investigate the mechanism of diquark condensate without this problem~\cite{Kogut}.

\section{Quark level Theory}
\subsection{Quark living in a complex representation}
In QCD, quarks live in the fundamental, a complex, representation. For general $n$ flavours of quarks, the Lagrangian with external sources can be written as
\begin{eqnarray}
\label{lagrangianQCD}
\mathcal{L} &=& \bar{q}_{L}i\gamma^\mu D_\mu q_{L} +
\bar{q}_{R}i\gamma^\mu D_\mu q_{R} \nn\\
&&\!\!\!+ \bar{q}_{L}\gamma^\mu l_\mu q_{L}+\bar{q}_{R}\gamma^\mu
r_\mu q_{R}\nn\\
&&-\bar{q}_{L}\mathcal{M^\dag} q_{R}-\bar{q}_{R}\mathcal{M}
q_{L}\,,
\end{eqnarray}
where $q_L$ and $q_R$ are column vectors and the
external fields $l_\mu$, $r_\mu$ and $\mathcal{M}=s-ip$ are matrices in flavour. The covariant derivative is given by $D_\mu = \partial_\mu q - iG_\mu q$.

When the external field vanish, this Lagrangian has $SU(n)_L\times SU(n)_R$  global flavour symmetry, which will be broken to $SU(n)_V$ spontaneously by the nonzero vacuum condensate $\langle\bar q q\rangle =\langle\bar{q}_L q_R + \bar{q}_R q_L \rangle \neq 0$. 

When the small quark mass term is present in the Lagrangian, the flavour symmetry $SU(n)_L\times SU(n)_R$ will be broken to $SU(n)_V$ explicitly, the Goldstone Boson (GB) gain mass and they become the pseudo-Goldstone Boson (PGB).

\subsection{Quarks living in a Real or pseudo-real representation}
Quarks can also live in real or pseudo-real representation in color space, their Lagrangian can be written as the following for two different cases.
\begin{itemize}
\item Quark live in real representation, e.g., adjoint representation.
\begin{eqnarray}
\label{lagrangianReal1}
\mathcal{L} &=&  \trc{\overline q_{Li}i\gamma^\mu D_\mu q_{Li}} +
           \trc{\overline q_{Ri}i\gamma^\mu D_\mu q_{Ri}}\nn\\&&\hspace{-0.4cm}
                 +\,\trc{\overline q_{Li}\gamma^\mu l_{\mu ij} q_{Lj}}
             +\trc{\overline q_{Ri}\gamma^\mu r_{\mu ij} q_{Rj}}\nn\\
&&\hspace{-0.4cm}-\,\trc{\overline q_{Ri}\mathcal{M}_{ij}q_{Lj}}
              -\trc{\overline q_{Li}\mathcal{M}^\dagger_{ij}q_{Rj}}\,,\nn\\
              D_\mu q &=& \partial_\mu q-iG_\mu q + i q G_\mu \,.
\end{eqnarray}
$\trc{A}$ means a trace over the gauge group indices and the quarks
are matrices rather than vectors in the gauge group indices. 
        
\item  Quark live in pseudo-real representation, e.g., two-color QCD.         
\begin{eqnarray}
\label{lagrangianPreal1}
\mathcal{L} &=&  \overline q_{Li}i\gamma^\mu D_\mu q_{Li} +
    \overline q_{Ri}i\gamma^\mu D_\mu q_{Ri}\nn\\&&
      +\overline q_{Li}\gamma^\mu l_{\mu ij} q_{Lj}
   +\overline q_{Ri}\gamma^\mu r_{\mu ij} q_{Rj}\nn\\
   &&-\overline q_{Ri}\mathcal{M}_{ij}q_{Lj}
   -\overline q_{Li}\mathcal{M}^\dagger_{ij}q_{Rj}\,,\\
D_\mu &=& \partial_\mu q - iG_\mu  q \nn \,,
\end{eqnarray}     
which has the same form of Lagrangian as QCD.
\end{itemize}   
One of the common features for those two theories is that they can have color singlet diquark and anti-diquark, which are the lightest baryons.

Using the following transformation, the left hand quark can be transfer to right hand anti-quark.
\begin{eqnarray}
\mathtt{Real:}&& \tilde q_{Ri} = C \overline q_{Li}^T\,,\\
\mathtt{Pseudo-real:}&& \tilde q_{R\alpha i} = \epsilon_{\alpha\beta} C \overline q_{L\beta i}^T\,,
\end{eqnarray}  
where the charge conjugate operator is $C=i\gamma^2\gamma^0$,  $\alpha, \beta$ are color indices,  and $\epsilon = i\sigma_2$. The $q_R$ and $\tilde q_R$ now transfer in the same way under chiral symmetry, so they can be placed in a same column vector
\begin{equation}
\hat q =\left(\begin{array}{c}q_R\\\tilde q_R\end{array}\right)\,.
\end{equation}  
Therefore the Lagrangian (\ref{lagrangianReal1}) and (\ref{lagrangianPreal1}) can be rewritten as the following,
\begin{itemize}
\item real representation
\begin{eqnarray}
\label{lagrangianReal2}
\mathcal{L} &=&
    \trc{\overline {\hat q} i\gamma^\mu D_\mu \hat q}
   +\trc{\overline {\hat q}\gamma^\mu \hat V_\mu \hat q_{j}}\\
   &&-\frac{1}{2}\trc{\overline{\hat q} C \hat \mathcal{M}\overline{\hat q}^T}
   -\frac{1}{2}\trc{ \hat q^T C \hat\mathcal{M}^\dagger \hat q}\nn\,,
\end{eqnarray}  
   
\item pseudo-real representation
\begin{eqnarray}
\label{lagrangianPreal2}
\mathcal{L} &=&    \overline {\hat q} i\gamma^\mu D_\mu \hat q
   +\overline {\hat q}\gamma^\mu \hat V_\mu \hat q_{Lj}\\
   &&
   -\frac{1}{2} \overline{\hat q}_\alpha C\epsilon_{\alpha\beta}
 \hat \mathcal{M}\overline{\hat q}^T_\beta
   -\frac{1}{2} \hat q_\alpha\epsilon_{\alpha\beta}
   C \hat\mathcal{M}^\dagger \hat q_\beta \nn\,.
\end{eqnarray}
\end{itemize} 
The external sources are now $2n\times 2n$ matrices,
\begin{eqnarray}
\hat V_\mu =  \left(\begin{array}{cc} r_\mu & 0 \\
     0 & \pm l_\mu^T \end{array}\right),
     \quad
\hat \mathcal{M} =
\left(\begin{array}{cc} 0 & \pm \mathcal{M}\\
\mathcal{M}^T & 0\end{array}\right)     
\end{eqnarray}  
where the `+' sign for real, the `-' sign for pseudo-real representation, respectively.

When the external fields vanish, the Lagrangian (\ref{lagrangianReal2}) and (\ref{lagrangianPreal2}) has $SU(2n)$ global symmetry rather than $SU(n)_L\times SU(n)_R$.

In the real case, the vacuum condensate $\langle\trc{ \bar q q}\rangle=\langle\trc{ \hat q^T C J_S \hat q}\rangle+ \mathrm{h.c.} \neq 0$ will spontaneously break $SU(2n)$ to $SO(2n)$, while in the pseudo-real case, the vacuum condensate is $\langle\trc{ \bar q q}\rangle=\langle \hat q_\alpha\epsilon_{\alpha\beta} C J_A \hat q_\beta\rangle
+ \mathrm{h.c.}\neq 0$, which breaks $SU(2n)$ to $Sp(2n)$ spontaneously. 
The $J_S$ and $J_A$ are symmetric and anti-symmetric $2n\times 2n$ matrices: 
\begin{eqnarray}
J_S=
\left(\begin{array}{cc} 0 & \mathbb{I}\\
\mathbb{I} & 0\end{array}\right),
\quad
J_A=
\left(\begin{array}{cc} 0 & -\mathbb{I}\\
\mathbb{I}& 0\end{array}\right)\,.
\end{eqnarray}
where $\mathbb{I}$ is the $n\times n$ unit matrix.

\section{Effective Field Theory}

\subsection{Goldstone Boson}
The Goldstone bosons (GB) live in the coset of broken symmetry $G/H$.
In terms of the pion fields $\phi^{a}$ , the nonlinearized matrix $u$  can be parametrized as
\begin{eqnarray}
u=\exp\left(\frac{i}{\sqrt{2}F_{0}}\sum_{a=1}^{N_{g}}\phi^{a}T^{a}\right)
\end{eqnarray} 
where $F_0$ is the bare decay constant of GB, $T^a$ is the generator of broken symmetry and normalized as $\trf {T^aT^b}=\delta^{ab}$.  The broken symmetry and number of GB $N_g$ for the three theories are summarized as the following,

\begin{itemize}
\item Complex representation:

$
G/H = SU(n)_L\times SU(n)_R /SU(n)_V, \quad N_g =n^2-1\,,
$
\item Real representation:

$
G/H = SU(2n)/SO(2n) ,\quad  N_g = n(2n+1)-1\,,
$
\item Pseudo-real representation:

$
G/H = SU(2n)/Sp(2n), \quad  N_g = n(2n-1)-1\,.
$
\end{itemize}

\subsection{The LO and NLO Lagrangian}
Using the method of CCWZ \cite{CCWZ}, all EFT of the three QCD-like theories can be written in a very similar way, which is the form of Chiral Perturbation Theory (ChPT). In the expansion of momentum, the leading order $\mathcal{O}(p^2)$ EFT Lagrangian is thus
\begin{equation}
\label{QCDLO}
\mathcal{L}_2 = \frac{F^2}{4}\langle u_\mu u^\mu +\chi_+\rangle\,.
\end{equation}
The NLO Lagrangian is \cite{GL1}
\begin{eqnarray}
\label{QCDNLO}
\mathcal{L}_4 &=&
L_0 \langle u^\mu u^\nu u_\mu u_\nu \rangle
+L_1 \langle u^\mu u_\mu\rangle\langle u^\nu u_\nu \rangle\nn\\&&
+L_2 \langle u^\mu u^\nu\rangle\langle u_\mu u_\nu \rangle
+L_3 \langle u^\mu u_\mu u^\nu u_\nu \rangle\nn\\&&
+L_4  \langle u^\mu u_\mu\rangle\langle\chi_+\rangle
+L_5  \langle u^\mu u_\mu\chi_+\rangle \nn\\&&
+L_6 \langle\chi_+\rangle^2
+L_7 \langle\chi_-\rangle^2
+\frac{1}{2} L_8 \langle\chi_+^2+\chi_-^2\rangle\nn\\&&
-i L_9\langle f_{+\mu\nu}u^\mu u^\nu\rangle
+\frac{1}{4}L_{10}\langle f_+^2-f_-^2\rangle\nn\\&&
+H_1\langle l_{\mu\nu}l^{\mu\nu}+r_{\mu\nu}r^{\mu\nu}\rangle
+H_2\langle\chi\chi^\dagger\rangle\,.
\end{eqnarray}
For the complex case, which is same as QCD, the objects appear in the Lagrangian are
\begin{eqnarray}
\label{QCDstandard}
u_\mu &=&i\left[u^\dagger(\partial_\mu-ir_\mu)u-u(\partial_\mu-l_\mu)u^\dagger\right]\,,\nn\\
\Gamma_\mu &=&\frac{1}{2}\left[u^\dagger(\partial_\mu-ir_\mu)u
+u(\partial_\mu-l_\mu)u^\dagger\right]\,,\nn\\
\chi_\pm &=& u^\dagger \chi u^\dagger\pm u\chi^\dagger u\,,\nn\\ 
f_{\pm\mu\nu}&=& u l_{\mu\nu} u^\dagger \pm u^\dagger r_{\mu\nu} u\,.
\end{eqnarray}
$l_{\mu\nu}$ and $r_{\mu\nu}$ are the field strengths from $l_\mu$ and $r_\mu$, $\chi$ include the mass matrix $\mathcal{M}$ via $\chi = 2B_0 \mathcal{M}$.

For real and pseudo-real case, the objects appear in the Lagrangian are
\begin{eqnarray}
\label{defumuadjoint}
u_\mu &=& 
i[u^\dagger(\partial_\mu-i \hat V_\mu)u - u(\partial_\mu+iJ \hat V^T_\mu J)u^\dagger]
\,,\nonumber\\
\Gamma_\mu &=& 
\frac{1}{2}[u^\dagger(\partial_\mu-i \hat V_\mu)u 
            + u(\partial_\mu+iJ\hat V^T_\mu J)u^\dagger]
\nonumber\,. \\
f_{\pm\mu\nu} &=&  
J u \hat V_{\mu\nu} u^\dagger J\pm  u \hat V_{\mu\nu} u^\dagger
\,, \nonumber\\
\chi_\pm &=& u^\dagger\hat\chi J u^{\dagger}  \pm  u J\hat\chi^\dagger u
\,.
\end{eqnarray}
where $J$ denotes the $J_S$ and $J_A$ in real and pseudo-real case, respectively. $\hat V_{\mu\nu}$ are the field strength of external fields, and $\hat\chi = 2B_0 \hat\mathcal{M}$
\cite{Bijnens:2009qm}. 

The form of NNLO Lagrangian for those three theories are also same as the ChPT case \cite{BCE1,BCE2} , which has 112+3 term.

One has to remember there are differences in different QCD-like theories, e.g., the generators, external sources, coupling constants, etc..

\subsection{The Renormalization}

When going to loop calculations, renormalization becomes
necessary. A thorough discussion of renormalization in ChPT
at NNLO can be found in \cite{BCE2,BCEGS}. For simplicity, the one loop divergences from $\mathcal{L}_2$ are absorbed by the bare coupling constant of $\mathcal{L}_4$, the two loop divergences from $\mathcal{L}_2$ and the one loop divergences from $\mathcal{L}_4$ are absorbed by the bare coupling constant of $\mathcal{L}_6$.
The divergence structure of NLO low energy constants (LECs) are
\begin{equation}
\label{defLir}
L_i = \left(c \mu\right)^{d-4}
\left[\Gamma_i\Lambda+L_i^r(\mu)\right]\,,
\end{equation}
with $\Lambda = 1/[16\pi^2(d-4)]$ and
$\ln c = -[\ln 4\pi+\Gamma^\prime(1)+1]/2$.
The constants $\Gamma_i$ are different for the three different theories. The complex case were calculated in \cite{GL2}. The pseudo-real case also been obtained in \cite{Splittorff}, which are slightly different from our results. The real case are obtained in \cite{Bijnens:2009qm}, where the typos are corrected in the table 1 of \cite{Bijnens:2011xt}.

The form of divergence structure of NNLO LECs $K_i$ is the same for three theories,
\ba
\label{defKir}
\hspace{-0.6cm} K_i = \left(c\mu\right)^{2(d-4)}
\left[K^r_i-\Gamma_i^{(2)}\Lambda^2
-\left(\frac{1}{16\pi^2}\Gamma_i^{(1)}+\Gamma_i^{(L)}
\right)\Lambda
\right]\,.
\ea
The coefficients $\Gamma_i^{(2)}$, $\Gamma_i^{(1)}$ and
$ \Gamma_i^{(L)}$ are only known for complex case \cite{BCE2}. We can still make sure the two loop calculation are correct by several methods, e.g., to check the cancellation of non-local divergences.

\section{The Calculation}

\subsection{Mass, vacuum condensate and decay constant}
The vacuum expectation value (VEV), mass and decay constant of of meson (PGB) were calculated up to NNLO in \cite{Bijnens:2009qm}. 

\vspace{0.2cm}
\begin{minipage}{0.9\linewidth}
\centering
\includegraphics[width=2.5in]{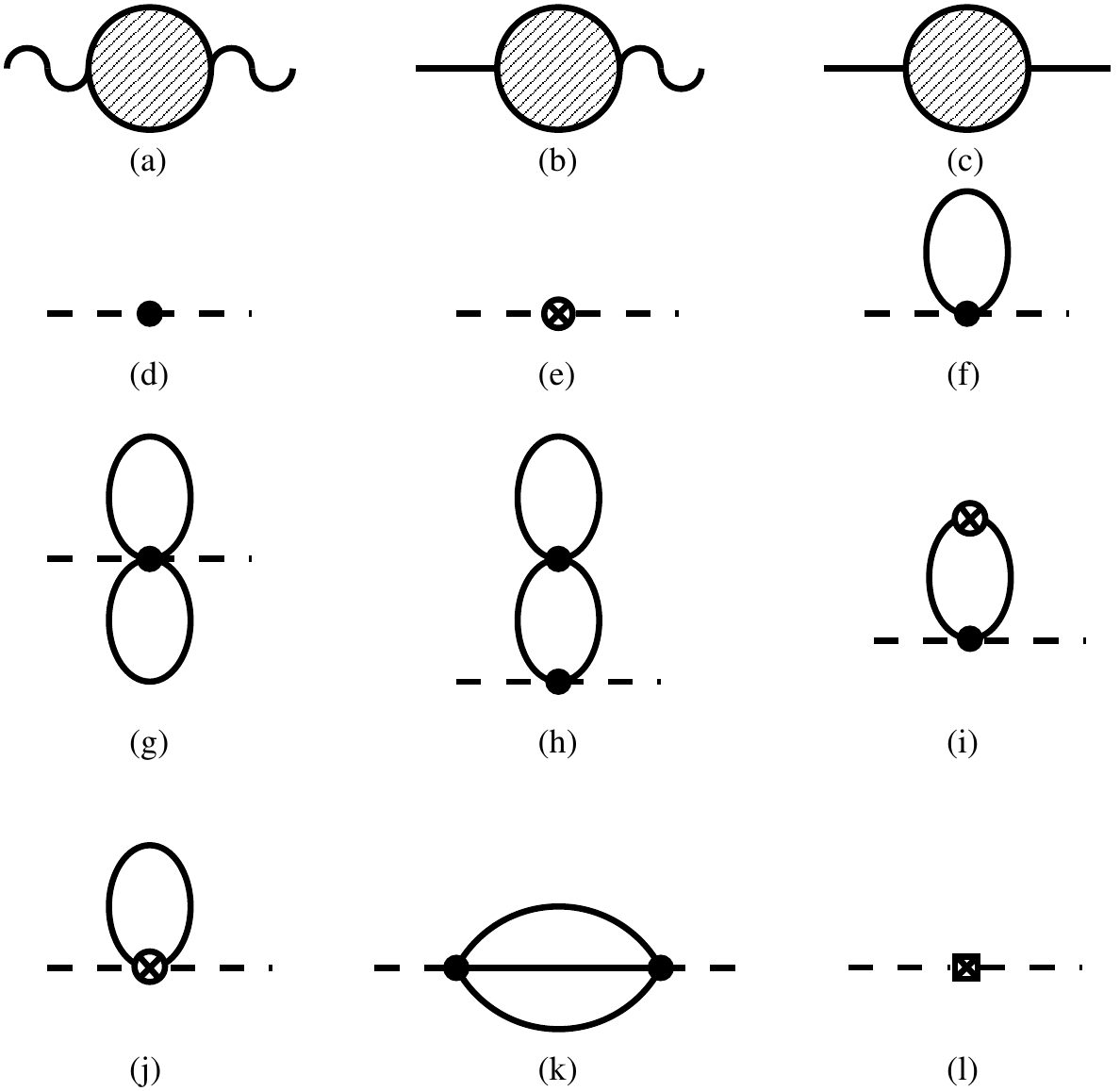}
\captionof{figure}{The 1PI diagrams up to order $p^6$. The wiggly lines indicate external sources, a solid line is a meson propagator, a wiggly line is an external
source, a dot is a vertex of $p^2$, a crossed circle is a vertex of $p^4$ and a crossed
box is a vertex of $p^6$.}
\label{1PI}
\end{minipage}
\vspace{0.2cm}

The mass can be obtained by finding the poles of the propagator, and the decay constant can be calculated by computing the one-meson matrix element of the axial current.
The 1PI diagrams up to NNLO are shown in Fig.\ref{1PI}.

We express the formula of these quantities in terms of the physical masses and decay constants
\begin{eqnarray}
O^2_\mathrm{phys} = O_\mathrm{LO}
+ O^2_\mathrm{NLO}+O^2_\mathrm{NNLO}\,,
\end{eqnarray}
where $O$ stands for $\langle \bar{q}q\rangle$, $M$ and $F$.
The $O_\mathrm{LO}$ only contains diagram (d), $O^2_\mathrm{NLO}$ contains diagrams (e) and (f), $O^2_\mathrm{NNLO}$ contains diagrams from (h) to (i). However, diagram (k) for VEV does not exist, since there is no trilinear interaction in the Lagrangian once we remove the external meson line from the IPI diagrams.
The details of our results can be found in \cite{Bijnens:2009qm}.  

\subsection{General meson-meson Scattering}
The general amplitude for meson-meson scattering $\phi^a \phi^b\to \phi^c \phi^d$ is given by
\begin{eqnarray}
\langle \phi^c(p_c)\phi^d(p_d)|\phi^a(p_a)\phi^b(p_d)\rangle
= M(s,t,u)\,.
\end{eqnarray}
The Mandelstam variables $s,t,u$ are defined by
\begin{eqnarray}
\hspace{-0.9cm} s = \frac{(p_a+p_b)^2}{\mphys}\,,\quad  t = \frac{(p_a-p_c)^2}{\mphys}\,,
\quad  u = \frac{(p_a-p_d)^2}{\mphys}\,.
\end{eqnarray}
According to different flavour structure of the scattering, the full amplitude can
be written in terms of two invariant amplitudes $B(s,t,u)$ and $C(s,t,u)$.
\begin{eqnarray}
\label{defBC}
&&\hspace{-1.0cm}M(s,t,u)\nn\\
\hspace{-1.0cm} &\hspace{-1.8cm}=& \hspace{-1.0cm} \left[\trf{X^a X^b X^c
X^d}+\trf{X^a X^d X^c X^b}\right] B(s,t,u)
\nonumber\\&&
\hspace{-1.2cm}+\left[\trf{X^a X^c X^d X^b}+\trf{X^a X^b X^d X^c}\right] B(t,u,s)
\nonumber\\&&
\hspace{-1.2cm}+\left[\trf{X^a X^d X^b X^c}+\trf{X^a X^c X^b X^d}\right] B(u,s,t)
\\&&
\hspace{-1.3cm}+\delta^{ab}\delta^{cd} C(s,t,u)+\delta^{ac}\delta^{bd} C(t,u,s)
+\delta^{ad}\delta^{bc} C(u,s,t)\,.\nn
\end{eqnarray}
The meson-meson scattering can be decomposed into many different scattering channels. In the language of group theory, this means that the direct product of two adjoint representations can be decomposed as direct sum of irreducible representations. For the $SU(n)$ or complex case, the  decomposition can be written as 
\begin{equation}
\label{decompositionSUN}
\hspace{-0.8cm}{Adj.}\otimes {Adj.} =
R_I \oplus R_S \oplus R_A \oplus+ R^{\ A}_S \oplus R^{\ S}_A
\oplus R^{\ A}_A \oplus R^{\ S}_S,
\end{equation}
where the subscript and superscript $S$ and $A$ denote `symmetric' and `anti-symmetric' for upper or lower indices, respectively. So there are 7 scattering channels for general $SU(n)$ case. This can be obtained Young tableaux or tensor method. In the case of $SO(2n)$ and $Sp(2n)$, the adjoint representation that contains mesons are symmetric or anti-symmetric, respectively. In both cases, there are 6 different channels
\begin{eqnarray}
&&\hspace{-1.5cm}\mathtt{Real:}\\
&&\hspace{-1.5cm} Sym.\otimes\!Sym. =
R_I \oplus R_A \oplus R_S \oplus R_{FS} \oplus R_{MA} \oplus R_{MS}\nn\,,\\
&&\hspace{-1.5cm}\mathtt{Pseudo-real:} \\
&&\hspace{-1.5cm} Asym.\otimes\!Asym. =
R_I \oplus R_A \oplus R_S \oplus R_{FA} \oplus R_{MA} \oplus R_{MS}\,,\nn
\end{eqnarray}  
where $FS$, $FA$ stands for `full symmetric' or `anti-symmetric' indices, $MA$, $MS$ stands for `mixed symmetric' or `anti-symmetric' indices, respectively.

Once the form of the representation is determined, the amplitude $T_r$ for different channels can be extracted from the general amplitude in two equivalent methods.
The first is to pick a state $R_r$ in a representation $r$ and get it via
\begin{equation}
T_r = \langle R_r|M(s,t,u)|R_r\rangle\,.
\end{equation}
The second is to apply the projection operators $P_r$ on the full amplitude $M(s,t,u)$ with
\begin{equation}
\label{stateamplitude}
P_r T_r = P_r M(s,t,u)\,.
\end{equation}
In the complex case, the amplitudes for each channel are

\begin{eqnarray}
\label{TISUN}
&&\hspace{-1.5cm}\mathtt{Complex:}\nn\\
T_I &=&  2\left(n-{1\over n}\right) [B(s,t,u) + B(t,u,s)] \nn \\
&&- {2\over n}B(u,s,t)
\nonumber\\
&& \hspace{-0.5cm} + (n^2-1) C(s,t,u) + C(t,u,s) +  C(u,s,t)\ ,
\nonumber\\
\hspace{-1.5cm}T_S &=&  \left(n - {4\over n}\right)[B(s,t,u) + B(t,u,s)]
\nonumber\\&&
- {4\over n}B(u,s,t)+ C(t,u,s) + C(u,s,t)\ ,
\nonumber\\
T_A &=& n[-B(s,t,u)+B(t,u,s)] \nn \\
&&+C(t,u,s)-C(u,s,t) \ ,
\nonumber\\
T_{SA} &=& C(t,u,s)- C(u,s,t) \ ,
\nonumber\\
T_{AS} &=& C(t,u,s) - C(u,s,t)\ ,
\nonumber\\
T_{SS}&=& 2B(u,s,t) + C(t,u,s) + C(u,s,t)\ ,
\nonumber\\
T_{AA}&=& \hspace{-0.2cm} -2B(u,s,t) + C(t,u,s) + C(u,s,t)\ .
\end{eqnarray}

For real case, the amplitudes for six channels are
\ba
\label{TISON}
&&\hspace{-1.5cm}\mathtt{Real:}\nn\\
T_I &=& {1\over n}(2n-1)(n+1)[B(s,t,u) + B(t,u,s)] \nn\\
&&  +{1\over n}(n-1)B(u,s,t) \nn\\
&&
\hspace{-1.2cm} +\,(2n-1)(n+1)C(s,t,u) + C(t,u,s) + C(u,s,t)\,,
\nonumber\\
T_A &=& - (1+n)[B(s,t,u) - B(t,u,s)] \nn \\
&&+ C(t,u,s) - C(u,s,t)\,,
\nonumber\\
T_S &=&{1\over n}(n-1)(n+2)[B(s,t,u)+B(t,u,s)] \nn\\
&& \hspace{-1.2cm} + {1\over n}(n-2)B(u,s,t) +C(t,u,s) + C(u,s,t)\,,
\nonumber\\
T_{FS} &=& 2B(u,s,t)+ C(t,u,s) + C(u,s,t)\,,
\nonumber\\
T_{MA} &=& C(t,u,s) - C(u,s,t)\,,
\nonumber\\
T_{MS} &=& \hspace{-0.2cm}-B(u,s,t)+C(t,u,s) + C(u,s,t)\,.
\ea
For pseudo-real case, the amplitudes are very similar to the real case,
\ba
\label{TISPN}
&&\hspace{-1.5cm}\mathtt{Pseudo-real:}\nn\\
T_I &=& {1\over n}(2n+1)(n-1)[B(s,t,u) + B(t,u,s)] \nn\\
&& -\, {1\over n}(n+1)B(u,s,t) \nn\\
&&
\hspace{-1.2cm} +\,(2n+1)(n-1)C(s,t,u) 
+ C(t,u,s) + C(u,s,t)\,,
\nonumber\\
T_A &=& {1\over n}(n+1)(n-2)[B(s,t,u)+B(t,u,s)] \nn\\
&&\hspace{-1.0cm} - {1\over n}(n+2)B(u,s,t) +C(t,u,s) + C(u,s,t)\,,
\nonumber\\
T_S &=& (1-n)[B(s,t,u) - B(t,u,s)] \nn\\
&&+ C(t,u,s) - C(u,s,t)\,,
\nonumber\\
T_{FA} &=& - 2B(u,s,t)+ C(t,u,s) + C(u,s,t)\,,
\nonumber\\
T_{MA} &=& C(t,u,s) - C(u,s,t)\,.
\nonumber\\
T_{MS} &=& B(u,s,t)+C(t,u,s) + C(u,s,t)\,.
\ea
They satisfy the relation
\be
\label{decompositionN}
M(s,t,u)=\sum_r T_r(s,t,u) P_r\ .
\ee

The scattering amplitude for each channel $I$
can be projected out using the partial wave expansion
\be
T^I_\ell(s) = {1\over64\pi} \int^1_{-1} d(cos\theta) P_\ell(cos\theta)
 T_I(s,t,u)\,.
\ee

Near the threshold $s=4$, we can expand the amplitude above the threshold
using $s=4(1+q^2/M^2_\pi)$ in the small three-momentum $q$.
\be
\mathrm{Re}\;T_\ell^I(s)=q^{2\ell}[a_\ell^I +q^2 b_\ell^I +O(q^4)]\,,
\label{eqthrex}
\ee
where $a_\ell^I$ is the scattering length, and $b_\ell^I$ is the slope.
The expressions of the lowest partial wave scattering length
for each channel in all three cases can be found in the appendix of \cite{Bijnens:2011fm}.

\subsection{Two point green function and S-parameter}
The definitions of the two-point functions are
\ba
\label{deftwop}
\hspace{-0.5cm} \Pi_{Va\mu\nu}(q) \hspace{-0.2cm} &\equiv & \hspace{-0.2cm}
i\int d^4x\; e^{iq\cdot x}\;\langle 0|T(V_\mu^a(x)V_\nu^a(0))^\dagger|0\rangle
\,,\nonumber\\
\Pi_{Aa\mu\nu}(q)\hspace{-0.2cm} &\equiv & \hspace{-0.2cm}
i\int d^4x\; e^{iq\cdot x}\;\langle 0|T(A_\mu^a(x)A_\nu^a(0))^\dagger|0\rangle
\,,\nonumber\\
\Pi_{Sa}(q)\hspace{-0.2cm} &\equiv & \hspace{-0.2cm}
i\int d^4x\; e^{iq\cdot x}\;\langle 0|T(S^a(x)S^a(0))^\dagger|0\rangle
\,,\nonumber\\
\Pi_{Pa}(q) \hspace{-0.2cm} &\equiv & \hspace{-0.2cm}
i\int d^4x\; e^{iq\cdot x}\;\langle 0|T(P^a(x)P^a(0))^\dagger|0\rangle 
\,, \qquad
\ea 
where $V_\mu^a$, $A_\mu^a$, $S^a$ and $P^a$ are vector, axial-vector, scalar and pseudo-scalar currents, respectively.

\vspace{0.2cm}
\begin{minipage}{0.9\linewidth}
\centering
\includegraphics[width=2.5in]{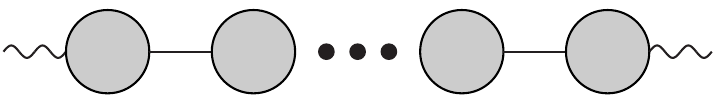}
\captionof{figure}{The illustrated Feynman diagram for two-point Green function. The filled circles indicate the 1PI diagrams, solid lines are
meson propagators and the wiggly lines indicate insertions of vector, axial-vector, scalar or pseudo-scalar current.
}
\label{twopoint}
\end{minipage}
\vspace{0.2cm}

Using Lorentz invariance the two-point functions with vectors and axial-vectors
can be decomposed in scalar functions
\be
\Pi_{Va\mu\nu} = (q_\mu q_\nu-q^2 g_{\mu\nu})\Pi_{Va}^{(1)}(q^2)
 +q_\mu q_\nu \Pi_{Va}^{(0)}(q^2)\;.
\ee
where $\Pi_{Va}^{(1)}(q^2)$ is the transverse part and
$\Pi_{Va}^{(0)}(q^2)$ is the longitudinal part or alternatively the spin 1 and
spin 0 part. The same definition holds
for the axial-vector two-point functions.

For a beyond the Standard Model with strong dynamics at the TeV scale,
there will in general be many resonances and other non-perturbative effects.
At low momenta one can use the EFT as described above for these cases.
We can estimate the Peskin-Takeuchi $S,T,U$ parameter \cite{Peskin:1991sw} contribution from pseudo-Goldstone Boson sector within the EFT. The parameter $T$ and $U$  vanish because of the exact flavour symmetry, i.e. we work in the equal mass case.
The $S$ parameter can be written as \cite{Peskin:1991sw}
\ba
S&=&-2\pi \Big[ \Pi'_{VV}(0) -\Pi'_{AA}(0) \Big]\nn \\
&=& 2\pi\frac{d}{dq^2}\left[q^2\Pi^{(1)}_{VV}-q^2\Pi^{(1)}_{AA}\right]_{q^2=0}\,.
\ea
$\Pi'_{VV}(0)$ and $\Pi'_{AA}(0)$ are the derivatives of the 
vector and axial-vector two-point functions at $q^2=0$.

The full results can be found in \cite{Bijnens:2011xt}, some plots for the purpose of illustration are also shown therein.

\section{Conclusion}
In this series of works, we have completed a comprehensive study of the Effective Field Theory of three QCD-like theories, which can be classified by having (techni-)quarks in complex, real or pseudo-real representations of the gauge group. They are corresponding to the spontaneously breaking of flavour symmetry, $SU(n)_L\times SU(n)_R \to SU(n)_V$, $SU(2n) \to SO(2n)$ and $SU(2n) \to Sp(2n)$, respectively.

Firstly, we constructed the effective theories for the three different cases in an extremely similar
way, obtain all the details of power counting and ready for high order calculation. Then
we calculated the vacuum condensate, mass and decay constant of meson up to NNLO
using the method of Chiral perturbation theory \cite{Bijnens:2009qm}. 

Secondly, we systematically
studied the general meson-meson scattering for those QCD-like theories. We constructed all the
possible intermediate states and scattering channels in the general $n$ flavour case, and
calculated the general amplitude and scattering lengths for each channel up to the NNLO \cite{Bijnens:2011fm}. 

We also calculated  vector, axial-vector, scalar and pseudo-scalar two-point Green functions up to NNLO. Using these results, we also estimated
the S-parameter contributing from pseudo-Goldstone-Boson  \cite{Bijnens:2011xt}.

Our results are useful for people working in the Lattice computing for strong dynamics at TeV scale, where the analytic formula for chiral extrapolation is needed. These results are also helpful for studying the diquark condensate mechanism at finite Baryon density.

\section*{Acknowledgements}
This work has been supported in part by the Spanish
Government and ERDF funds from the EU Commission
[Grants No. FPA2011-23778, No. CSD2007-00042
(Consolider Project CPAN)] and by Generalitat
Valenciana under Grant No. PROMETEOII/2013/007.
J. Lu is grateful to the hospitality of Center for Future High Energy Physics (CFHEP) in Beijing while this proceeding was written.




\nocite{*}
\bibliographystyle{elsarticle-num}
\bibliography{martin}



\end{document}